\documentstyle[prb,aps,twocolumn,epsf]{revtex}
\def\geqap{\,\raise 2pt \hbox{$>\kern-11pt \lower 5pt \hbox{$\sim$}$}\,}
\def\leqap{\,\raise 2pt \hbox{$<\kern-10pt \lower 5pt \hbox{$\sim$}$}\,}
\begin{document}
\draft
\title{Theory of spin wave excitation in manganites}
\author{Ryo Maezono and  Naoto Nagaosa}
\address{Department of Applied Physics, University of Tokyo,
Bunkyo-ku, Tokyo 113-8656, Japan}
\date{\today}
\maketitle
\begin{abstract}
\par
The role of the orbital degrees of freedom is studied theoretically
for the spin dynamics of $R_{1-x}A_x$MnO$_3$.
Based on the meanfield solution, an RPA calculation has been done
and it is found that the $d_{x^2-y^2}$-type orbital is essential 
for the double-exchange (DE) interactions, i.e., the DE is basically 
two-dimensional interaction.
Based on this results compared with experiments, we propose
that the orbital wavefunction is $d_{x^2-y^2}$-type
locally even in the metallic ferromagnetic state, which fluctuate 
quantum mechanically.
Well agreement of the estimation with experiments suggest that
the Jahn-Teller phonon has less importance on the spin dynamics.
\end{abstract}
%
\narrowtext
Doped manganites $R_{1-x}A_x$MnO$_3$ ($R$=La, Pr, Nd, Sm ;
$A$= Ca, Sr, Ba) have recently attracted considerable interests 
due to the colossal magnetoresistance (CMR) observed near the
ferromagnetic (spin $F$-type) transition temperature $T_c$ 
\cite{chaha,hel,LaSr1,jin}.
It is now recognized that the most fundamental interaction in 
these materials is the double exchange-interaction (DE),
which connects the transport and magnetism \cite{degennes}. 
Therefore the magnetism is a key issue to reveal the mechanism of CMR.
Especially, rich magnetic phase diagrams have been clarified 
over the wide range of the concentration $x$ and also the bandwidth.
With increasing $x$, the parent insulator with a layered 
antiferromagnetism (spin $A$-type AF) changes into 
a ferromagnetic metal (FM). \cite{jonker}
In addition to these well-known magnetic phases,
spin $A$-type AF (in La$_{1-x}$Sr$_{x}$MnO$_3$, \cite{Moritomo1} 
Pr$_{1-x}$Sr$_{x}$MnO$_3$, \cite{kawano97,PrSr}
Nd$_{1-x}$Sr$_{x}$MnO$_3$ \cite{Kuwahara1}) and
the rod type antiferromagnetism (spin $C$-type AF, 
in Nd$_{1-x}$Sr$_{x}$MnO$_3$ \cite{Kuwahara1}) were recently found 
in the moderately doped metallic region ($0.5<x<0.8$).
\par
The neutron scattering experiments have revealed the
spin wave excitation at low temperatures, which depends sensitively on 
the doping $x$ and the magnetic structure 
\cite{martin,perring,endoh1,hirota1,hirota2,fernandez,yoshizawa,hwang1}.
In the spin $A$-type AF for small $x$, the dispersion is two dimensional while
it becomes isotropic in FM state. 
The spin stiffness, however, stays almost constant up to 
$x \cong 0.125$ where the phase transition between the 
insulating and metallic ferromagnetic states occurs. \cite{endoh1} 
This phase transition is accompanied with that of the orbital 
structure. \cite{endoh99}
In the FM state, the orbital ordering disappears and the spin stiffness 
begins to increase.
In the metallic $A$-type AF (AFM) state for higher $x$, the dispersion
becomes again two dimensional.\cite{yoshizawa}
In this paper we report the calculation of the spin wave dispersion
by changing $x$ and taking into account the orbital structure.
The calculated $x$-dependence of the spin stiffness agrees quantitatively
with the observed one for $x \gtrsim 0.2$ where the double-exchange 
interaction dominates.
This $x$-dependence strongly supports the large orbital polarization,
which is $d_{x^2-y^2}$ at least locally.
Therefore the double-exchange interaction is basically two-dimensional.
\par
We previously reported a meanfield theory (MFT)
for the phase diagram of doped manganites in terms of a model including 
the strong on-site repulsion, orbital degeneracy, and anisotropic
covalency \cite{maezono2}. 
Based on this MFT, we first 
presents the spin wave dispersion in terms of the 
random-phase-approximation (RPA).
This reproduces qualitatively the $x$-dependence of the stiffness
and the anisotropy
due to the cross-over from superexchange interaction (SE) to DE.
Especially for the doped region, only when the orbital configuration 
becomes $d_{x^2-y^2}$, i.e., $x \gtrsim 0.2$, the DE becomes 
appreciable and the in-plane spin stiffness grows 
rapidly.
Observed values of the in-plane spin stiffness
\cite{martin,fernandez,yoshizawa} 
agree  quantitatively with the estimated value with 
$d_{x^2-y^2}$-orbital ordering.
This is understood in terms of the orbital liquid picture 
\cite{maezono2,ishihara2} and implies that the Jahn-Teller (JT) phonon 
has less importance on the spin dynamics.
\par
We start with the Hamiltonian
\begin{eqnarray}
H&=&
\sum\limits_{\sigma \gamma \gamma' \langle ij \rangle} 
{t_{ij}^{\gamma \gamma '}d_{i\sigma \gamma }^{\dagger}d_{j\sigma \gamma '}}
\nonumber \\
&-&
J_H\sum\limits_i {\vec S_{t_{2g} i}\!\cdot\! 
\vec S_{e_g i}}
\nonumber \\
&+& J_S\sum\limits_{\left\langle {ij} \right\rangle } {\vec S_{t_{2g} i}
\!\cdot\! \vec S_{t_{2g} j}} +H_{\rm on\ site}
\label{eqn:eq1}
\end{eqnarray}
where $\gamma$ [$=a(d_{x^2-y^2}), b(d_{3z^2-r^2})$] specifies the 
orbital and the other notations are standard.
The transfer integral $t_{ij}^{\gamma \gamma'}$ depends on the pair 
of orbitals $(\gamma, \gamma')$ and the direction of the 
bond $(i, j)$ \cite{ishihara2}.
The spin operator for the $e_g$ electron is defined as 
$\vec S_{e_g i}={1 \over 2}\sum\limits_{\gamma \alpha \beta} 
 {d_{i\gamma \alpha }^{\dagger}\vec \sigma _{\alpha \beta }
d_{i\gamma \beta }}$ with the Pauli matrices $\vec \sigma$, 
while  the orbital isospin operator is defined as 
$\vec T_i={1 \over 2}\sum\limits_{\gamma \gamma' \sigma}  {d_{i\gamma \sigma }
^\dagger\vec \sigma _{\gamma \gamma '}d_{i\gamma '\sigma }} \ .$
$J_H$ is the Hund's coupling between $e_g$ and 
$t_{2g}$ spins, and $J_S$ is 
the $AF$ coupling between nearest neighboring $t_{2g}$ spins.
$H_{\rm on\ site}$ represents the on-site
Coulomb interactions between $e_g$ electrons. Coulomb interactions
induce both the spin and orbital isospin moments, and actually
$ H_{\rm on\ site} $ can be written as 
\cite{maezono2,ishihara2}
\begin{equation}
H_{\rm on\ site}
= -\sum\limits_i {\left( {\tilde \beta \vec T_i^2+\tilde
 \alpha \vec S_{e_{g} i}^2} \right)}, 
\label{eqn : eq2}
\end{equation}
where the coefficients of the spin and isospin operators, i.e., 
$\tilde \alpha $ and $\tilde \beta $, are given by \cite{maezono2,ishihara2}
$\tilde \alpha = U-{J \over 2}>0\ ,$
and $\tilde \beta = U-{3J \over 2}>0 \ .$
The parameters $\tilde \alpha ,\tilde \beta ,
t_0$, used in the numerical calculation are chosen as
$t_0 \sim 0.72$ eV, $U=6.3$ eV, and $J=1.0$ eV, being 
relevant to the actual manganites. \cite{maezono2} 
\par
In the path-integral quantization, we introduce the Stratonovich-Hubbard 
fields $\vec \varphi_S$ and $\vec \varphi_T$, representing the spin and 
orbital fluctuations, respectively.
With the large values of the electron-electron interactions above, 
both $\vec \varphi_S$ and $\vec \varphi_T$ are almost fully polarized.
\cite{maezono2}
The MFT corresponds to the saddle point configuration 
of $\vec \varphi_S$ and $\vec \varphi_T$.
We consider four kinds of spin alignment in the cubic cell: $F$-, 
$A$-, $C$- and $G$-type. 
As for the orbital degrees of freedom, we consider two sublattices $I$,
and $I\!I$, on each of which the orbital is specified by the
angle $\theta_{I,I\!I}$ as \cite{maezono2}
\begin{eqnarray}
\left| {\theta _{I,I\!I}} \right\rangle =\cos {{\theta _{I,I\!I}} 
\over 2}\left| {d_{x^2-y^2}} \right\rangle +\sin {{\theta _{I,I\!I}} 
\over 2}\left| {d_{3z^2-r^2}} \right\rangle.
\label{eqn : eqN.3}    
\end{eqnarray}
We consider four types, i.e., 
$F$-, $A$-, $C$-, $G$-type also for the orbital ordering.
Henceforth, we  use a notation such as spin A, orbital $G$
($\theta_I,\theta_{I\!I}$) etc..
In MFT, the most stable ordering is given by
\cite{maezono2}
\begin{tabbing}
xxx \= $x=0.5-0.999$  \= Spin CCC \=  Orbital F:(180) \kill
\>$x=0.0$ \> Spin A \> Orbital C:($60,-60$) \\
\>$x=0.1$ \> Spin F \> Orbital C:($80,-80$) \\
\>$x=0.2-0.4$ \> Spin A \> Orbital F:($0$,$0$) \\
\>$x=0.5-0.9$ \> Spin C \> Orbital F:($180$,$180$). \\
\label{tbl : saddle point}
\end{tabbing}
\noindent
As for $x=0$, we further introduced the JT effect \cite{maezono2} by putting
the observed distortion of the MnO$_6$ octahedra.\cite{gen}
\par
RPA corresponds to the Gaussian fluctuation around MFT, and the 
contribution to the spin wave effective action from the $e_g$-electrons
$S_{\rm SW}$ is obtained as the expansion around the saddle point.
\begin{eqnarray}
\lefteqn{
S_{\rm SW}=
\sum\limits_{q,\Omega } {K_\pi \left( {\vec q,\Omega } \right)
 \pi \left( {\vec q_S\!+\! \vec q,\Omega } \right)
\!\cdot\!
\pi }
\left( {-\vec q_S\!-\!\vec q,-\Omega } \right)
} \cr
&\ \ \ \ & \!+\!\sum\limits_{q,\Omega }\! {K_{\!\times\!}\! 
\left( {\vec q,\Omega } \right) \vec \pi \!
\left( {\vec q_S\!+\!\vec q,\Omega } \right)
\!\cdot\! \left\{ {\vec n\!\times\! \vec \pi 
\left( {-\vec q_S\!-\!\vec q,-\Omega } \right)} \right\}}.
\label{eqn : eq3}    
\end{eqnarray}
where 
$\vec q_S(\equiv -\vec q_S)$ is the wavevector and 
$\vec n$ $(| \vec n | = 1)$ is the direction of the 
ordered magnetic moment, and $\vec \pi$ is the 
fluctuation perpendicular to it.
Because the spin wave is the Goldstone boson, the condition
$
K_\pi \left( {0,0} \right)=0 \ , 
K_{\!\times\!} \left( {0,0} \right)= 0 ,
$
can be derived.
Coefficient of the diagonalized quadratic form is obtained as
$K_{\uparrow \left( \downarrow  \right)}
=K_\pi \pm iK_\times \ ,$
zero-point of which $\left(K_{\uparrow \left( \downarrow  \right)}
(\vec q,\Omega=-i\omega)=0\right)$ gives the dispersion relation of 
the excitation $\omega=\omega(\vec q)$. 
However in this paper we focus on the {\it static} spin stiffness
rather than the {\it dynamic} spin wave velocity because 
(a) at $x=0$ the spin stiffness is correctly reproduced to be
of the order of $J$ in the RPA while the spin wave velocity scales with $t$,
and (b) for the metallic region, $x\ne 0$, 
the Landau-damping is not properly treated in our calculation where
the Brillouin zone is discretized and thus the gapless 
individual-excitation is not correctly evaluated.
The {\it static} spin stiffness $C_{\alpha }$
corresponds to the static response function for small $|\vec q|$ as
\begin{equation}
{{K_{\sigma}\left( {\vec q,\Omega=0 } \right)} \over {\tilde \alpha }}
\cong \sum\limits_{\alpha =x,y,z} 
{C_\alpha q_\alpha ^2}
\ ,
\label{eqn : eqN.1}
\end{equation}
and roughly reflects the exchange-interaction depending on $x$,
where $\sigma=1\ (-1)$ corresponds to spin up (down), respectively. 
\par
In RPA calculation the SE corresponds to the contribution
from the inter-band transitions, while the DE from the 
intra-band ones.
In this way, the present calculation describes both SE and DE interactions, 
and hence their crossover in a unified way.
Also the contribution from $J_S$ should be considered, the value of 
which is determined in the following way.
We require that the experimentally observed anisotropy ratio 
of the spin stiffness
$R = \left( {{{D_{x,y}} \over {D_z}}} \right)^2$ 
is reproduced when the calculated contributions
from $e_g$-electrons and that from $J_S$ are added.
The observed value $R=7.6$ for LaMnO$_3$ \cite{hirota2}
leads to an estimation as $J_S=0.997$ meV.
As for AFM at $x=0.3$, the observed value $R=10.4$
for Nd$_{0.45}$Sr$_{0.55}$MnO$_3$ \cite{yoshizawa}
gives $J_S=1.4$ meV.
These estimations are consistent rather with $J_S\sim 0.8$ meV
estimated from the N\'eel temperature of CaMnO$_3$ \cite{goodenough} 
than the earlier meanfield estimations $J_S\sim8$ meV.
\cite{maezono2,ishihara96}
Using these estimations, $J_S\sim 1$ meV, we can estimate the 
spin wave stiffness for $x=0$ as
$J_{\rm total}^{x} S^2_{\rm total} = 1.05$ meV,
including the contribution from $t_{2g}$ orbital.
The corresponding experimental value is
$J_{\rm total}^{x} S^2_{\rm total} = 3.91$ meV 
in LaMnO$_3$ \cite{hirota2} with the reported lattice constants and the 
magnitude of spin moment, $S_{\rm total}=3/2+1/2(1-x)$.
The discrepancy may be attributed to the complex lattice deformations
such as the Mn-O bond-length (JT-type distortion) and the Mn-O-Mn 
bond-angle (orthorhombic distortion) observed at $x=0$ \cite{hirota2}, 
which can also be an origin of the anisotropy \cite{igor,sawada}.
\par
\begin{figure}[p]
\begin{center}
\vspace{0mm}
\hspace{0mm}
\epsfxsize=8cm
\epsfbox{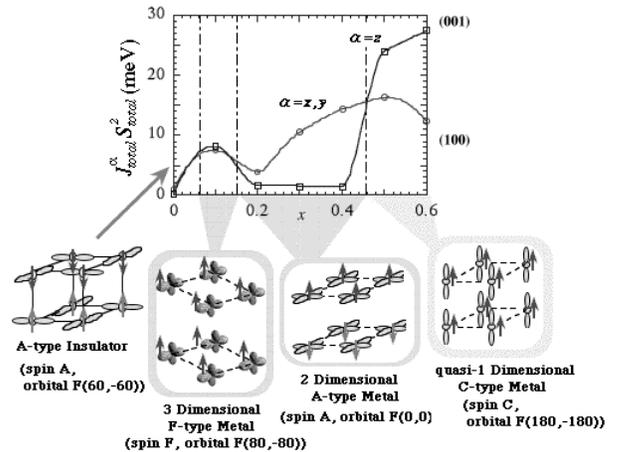}
\vspace{0mm}
\caption[aaa]{Doping-dependence of the spin stiffness.
The orbital and the spin structure is optimized at each point.
The enhancement of the spin-stiffness and the cross-over of 
the dimensionality are seen with increasing $x$.}
\label{fig : B2}
\end{center}
\end{figure}
\noindent
\par
We now turn to the doped case $x \ne 0$. 
Fig. \ref{fig : B2} shows the doping-dependence of the total stiffness
calculated for the optimized spin/orbital structure at each $x$.
Firstly the spin stiffness due to the 
double exchange interaction scales roughly with
$x$ because the orbital
is almost fully polarized, while in the absence of the orbital polarization
it scales with the electron density $(1-x)$ rather than the hole $x$ for 
small $x$.
The observed stiffness enhancement with increasing $x$ even in the metallic
region \cite{endoh1}
therefore also supports the large orbital polarization due to the
strong Coulomb interactions.
As $x$ increases, the spin structure changes from
spin $A$-type insulator at $x=0$ into the nearly isotropic FM,
to the AFM with two-dimensional $d_{x^2-y^2}$
orbital alignment, and to the spin $C$ metal with $d_{3z^2-r^2}$
orbital. \cite{maezono2}
Accordingly, the in-plane stiffness shows an increase,
moderately at the beginning and then rapidly in the region of AFM. 
This reflects the fact that the DE is the most 
effective and prefers the $d_{x^2-y^2}$-orbital, i.e., the DE
is basically {\it two-dimensional} with the $e_g$-orbitals.
In the spin-$C$-metal for $x>0.4$, one-dimensional orbital along
(001)-direction gives rise to a steep increase of the stiffness
in this direction.
\par
The observed anisotropy of the spin stiffness is 
determined by the long range ordering of the orbitals.
Fig. \ref{fig : B2} also represents the cross-over of the 
dimensionality which we proposed in the previous report.
\cite{maezono2}
Yoshizawa $et\ al.$ \cite{yoshizawa} observed the reentrant of
such two-dimensional anisotropy of the stiffness for 
Nd$_{0.45}$Sr$_{0.55}$MnO$_3$, being consistent with our result.
Quasi-one-dimensional anisotropy is predicted
for Nd$_{1-x}$Sr$_x$MnO$_3$ ($x>0.6$) \cite{Kuwahara1}.
\par
The in-plane spin stiffness $J_{\rm total}^{x\left(y\right)} S^2_{\rm total}$
in Fig. \ref{fig : B2} could be compared with the experiments.
In La$_{1-x}$Sr$_x$MnO$_3$, Endoh $et\ al.$\cite{endoh1}
observed the plateau of the velocity $v_x$ in the orbital-ordered
insulating state up to $x\sim0.12$ and then the velocity increases
in the FM phase.
Comparing this with the calculation above, it seems that the 
moderate increase up to $x\sim0.15$
in Fig. \ref{fig : B2} corresponds to the plateau, while
the rapid increase for $x>0.15$ to the increasing velocity
observed by Endoh.\cite{endoh1}
Then orbital-ordered FM state in Fig. 1
corresponds to the insulating spin $F$ phase in experiments.
Both the FM and AFM phases in experiments, on the other hand,
seems to corresponds to the AFM with $d_{x^2-y^2}$ orbital ordering
in the calculation.
This fits well orbital liquid picture by Ishihara $et\ al.$
\cite{ishihara2};
In a perfectly cubic system the orbital state in FM is described as
the resonance among $d_{x^2-y^2}$, $d_{y^2-z^2}$, and $d_{z^2-x^2}$.
In the actual CMR compounds, however, the slight lattice distortion 
\cite{martin,fernandez} may breaks the cubic
symmetry to stabilize $d_{x^2-y^2}$ though it is still accompanied with large 
fluctuation around it.
\par
%
Now we turn to the absolute value of the spin stiffness in FM phase.
With the reported lattice constants the experimental values of the 
spin stiffness, $J_{\rm total}^{x} S^2_{\rm total}$, are
11.61 meV for La$_{0.7}$Sr$_{0.3}$MnO$_3$ \cite{martin}
and 10.24 meV for Nd$_{0.7}$Sr$_{0.3}$MnO$_3$ \cite{fernandez}, respectively.
These are in a good agreement with $J_{\rm total}^x S^2_{\rm total}$
=10.53 meV  estimated by RPA here with $x=0.3$, $d_{x^2-y^2}$-orbital ordering
(a simple tight binding estimation with $d_{x^2-y^2}$-orbital also gives 
similar value).
This agreement implies the large orbital polarization in FM phase
with $d_{x^2-y^2}$ at least locally.
\par
This orbital liquid picture also explains the spin wave softening
\cite{fernandez,hwang1,dai99} and spin canting 
\cite{kawano98} observed in this system.
Some theoretical works shows that the orbital fluctuation such an orbital
liquid state leads to the softening of the spin wave dispersion near the 
zone boundary \cite{khaliullin99} with the anisotropic feature 
\cite{nagaosa99} (the softening almost disappears along
$(\pi,\pi,\pi)$-directions \cite{hwang1,dai99}).
As for the spin canting,
the observed canting in the metallic region (Nd$_{0.5}$Sr$_{0.5}$MnO$_3$)
with the FM/AFM transition \cite{kawano98} cannot be explained unless 
the planer orbital $d_{x^2-y^2}$ realizes in FM phase \cite{maezono99}
(observed slight lattice anisotropy can not stabilize such a 
planer orbital without the occurrence of the orbital liquid state).
\par
%
An important conclusion from the agreement
between the experiments and RPA calculation of the stiffness constant
is that the polaron effect is small in the metallic state 
\cite{millis95} at least on the spin dynamics.
Polaron should reduce the DE interaction in the doped region 
via a bandwidth reduction by a factor of 
$< X^\dagger_i X_j > = \exp[ - \sum_q | u_q |^2/2]$
($ u_q = (g_q/\omega_q)
( e^{i q \cdot R_i } -  e^{i q \cdot R_j } )$),
where 
$X_i = \exp[ \sum_q e^{ i q \cdot R_i} 
( g_q/\omega_q)(b_q - b^\dagger_{-q})]$
is a factor encountered in the canonical transformation eliminating
the coupling between electrons and polaronic bosons,
$\sum_{i,\sigma} \sum_q g_q ( b_q + b^\dagger_{-q}) 
d^\dagger_{i \sigma} d_{i \sigma}$, with the coupling constant
$g_q$ and phonon frequency $\omega_q$.\cite{mahan}
On the other hand, for $x=0$, the SE under the coupling with the polaron 
is given by,
\begin{equation}
J \!=\! 4 |t_{ij}|^2 \! \int_0^\beta \! d \tau G_0^2(\tau)\! 
\left<
\! X^\dagger_i(\tau) X_j(\tau) X^\dagger_j(0) X_i(0) \!\right>
\ ,
\end{equation}
where $G_0(\tau) = e^{- U \tau/2}$ is the Green's function 
for localized electrons.
Because we are interested in the large $U$ case, 
the integral is determined by the small $\tau$ region where
$\left< \! X^\dagger_i(\tau) X_j(\tau) X^\dagger_j(0) X_i(0) \!\right>
\cong e^{ - { \tilde \Delta} \tau }$ 
($ { \tilde \Delta } = \sum_q \omega_q |u_q|^2 $).
Then the polaronic effect is to replace $U$ by $U + {\tilde \Delta}$
in the expression for $J$, which is a minor correction when 
$U>> {\tilde \Delta}$ \cite{kugel}, being in sharp contrast to DE 
discussed above.
Polaronic effect should therefore correct the RPA-estimation of the
stiffness-enhancement as $x$ increases to be smaller.
Agreement between the observed and estimated stiffness for DE
implies therefore that the spin dynamics is not so affected by
the polaron.
This is also pointed out by Quijada \it{et al}\rm.\cite{quijada}
\par
%
In summary, we have studied the role of orbitals 
in the spin dynamics of $R_{1-x}A_x$MnO$_3$.
Comparing the experiments with the RPA calculation based on the mean
field theory, we conclude the followings.
(a) $x$-dependence of the stiffness-enhancement suggests the large orbital
polarization.
(b) the double-exchange interaction prefers $d_{x^2-y^2}$ orbital and 
is basically two-dimensional interaction, which leads to the large 
anisotropy of the spin dynamics.
(c) the agreement between experiments and RPA results strongly suggests 
that the spin dynamics is not so affected by the JT polaron.
\par
 The authors would like to thank K. Hirota, Y. Endoh, I. Solovyev,
K. Terakura, R. Kajimoto, H. Yoshizawa, T. Kimira, D. Khomskii, 
A. Millis, and Y. Tokura for their valuable discussions.
 This work was supported by Priority Areas Grants from the Ministry 
of Education, Science and Culture of Japan.
\par
%
%


\begin{references}
%
\bibitem{chaha}
K. Chahara \it{et al.}\rm,
{\bf 62}, 780 (1993).
%
\bibitem{hel}
R. von Helmolt \it{et al.}\rm,
Rev. Lett. {\bf 71}, 2331 (1993). 
%
\bibitem{LaSr1}
A. Urushibara \it{et al.}\rm, Phys. Rev. B {\bf 51}, 14103 (1995). 
%
\bibitem{jin}
S. Jin \it{et al.}\rm, Science, {\bf 264}, 413 (1994). 
%
\bibitem{degennes}
P. G. de Gennes, Phys. Rev.  {\bf 118}, 141 (1960). 
%
\bibitem{jonker}
 G. H. Jonker, and H. van Santen, Physica {\bf 16}, 337 (1950).
%
\bibitem{Moritomo1}
 Y. Moritomo \it{et al.}\rm, Phys. Rev. B {\bf 58}, 5544 (1998).
%
\bibitem{kawano97}
H. Kawano \it{et al.}\rm, Phys. Rev. Lett. {\bf 78}, 4253 (1997).
%
\bibitem{PrSr}
Y. Tomioka \it{et al.}\rm, (unpublished).
%
\bibitem{Kuwahara1}
H. Kuwahara \it{et al.}\rm, Mat. Res. Soc. Sym. Proc. {\bf 494}, 83 (1998).
%
\bibitem{martin}
M.C. Martin \it{et al.}\rm, Phys. Rev. B {\bf 53}, R14285 (1996).
%
\bibitem{perring}
T. G. Perring \it{et al.}\rm, Phys. Rev. Lett. {\bf 77}, 711 (1996).
%
\bibitem{endoh1}
Y. Endoh \it{et al.}\rm, J. Phys. Soc. Jpn. {\bf 66}, 2264 (1997).
%
\bibitem{hirota1}
K. Hirota \it{et al.}\rm, Physica B {\bf 36}, 237 (1997).
%
\bibitem{hirota2}
K. Hirota \it{et al.}\rm, J. Phys. Soc. Jpn. {\bf 65}, 3736 (1996).
%
\bibitem{fernandez}
J. A. Fernandez-Baca \it{et al.}\rm, Phys. Rev. Lett. {\bf 80}, 4012 (1998).
%
\bibitem{yoshizawa}
H. Yoshizawa \it{et al.}\rm, Phys. Rev. B {\bf 58}, R571 (1998).
%
\bibitem{hwang1}
H. Y. Hwang \it{et al.}\rm, Phys. Rev. Lett. {\bf 80}, 1316 (1998).
%
\bibitem{endoh99}
Y. Endoh \it{et al.}\rm, Phys. Rev. Lett. {\bf 82}, 4328 (1999).
%
\bibitem{maezono2}
R. Maezono \it{et al.}\rm, Phys. Rev. B {\bf 57}, R13993 (1998),
Phys. Rev. B {\bf 58}, 11583 (1998).
%
\bibitem{ishihara2}
S. Ishihara \it{et al.}\rm, Phys. Rev. B {\bf 56}, 686 (1997). 
%
\bibitem{gen}
G. Matsumoto, J. Phys. Soc. Jpn. {\bf 29}, 606 (1970).
%
\bibitem{goodenough}
J. B. Goodenough, Phys. Rev. {\bf 100}, 564 (1955).
%
\bibitem{ishihara96}
S. Ishihara \it{et al.}\rm, Phys. Rev. B {\bf 55}, 8280 (1997). 
%
\bibitem{igor}
I. Solovyev \it{et al.}\rm, Phys. Rev. Lett. {\bf 76}, 4825 (1996).
%
\bibitem{sawada}
H. Sawada \it{et al.}\rm, Phys. Rev. B {\bf 56}, 12154 (1997). 
%
\bibitem{khaliullin99}
G. Khaliullin and R. Kilian, preprint (cond-mat/9904316).
%
\bibitem{nagaosa99}
 N. Nagaosa, in preparation.
%
\bibitem{dai99}
 P. Dai \it{et al.}\rm, preprint (cond-mat/9904372).
%
\bibitem{kawano98}
H. Kawano \it{et al.}\rm, preprint (cond-mat/9808286).
%
\bibitem{maezono99}
R. Maezono \it{et al.}\rm, preprint (cond-mat/9904427).
%
\bibitem{millis95}
 A.J. Millis \it{et al.}\rm, Phys. Rev. Lett. {\bf 74}, 5144 (1995).
%
\bibitem{mahan}
G.D. Mahan, in \it Many-Particle Physics, 2nd ed., 
\rm Chap. 4 (Plenum Press, New York, 1990). 
%
\bibitem{kugel}
K.I. Kugel \it{et al.}\rm, Sov. Phys. JETP {\bf 52}(3), 501 (1981).
%
\bibitem{quijada}
M. Quijada \it{et al.}\rm, Phys. Rev. B {\bf 58}, 16093 (1998). 
%
%
\end{references}
\end{document}